\documentclass[journal=jctcce,manuscript=article]{achemso}

\usepackage{amsmath}
\usepackage{bm}% bold math
\usepackage[utf8]{inputenc}
\usepackage{tabularx}
\usepackage{graphicx}% Include figure files
\usepackage{dsfont}
\usepackage{dcolumn}% Align table columns on decimal point
\usepackage{lipsum}
\usepackage{color}
\usepackage{braket}
\usepackage{float}
\usepackage{multirow}
\usepackage{subfig}
\usepackage{makecell}

\title{Hardware Efficient Quantum Algorithms for Vibrational Structure Calculations}
\author{Pauline J. Ollitrault}
\affiliation{IBM Research GmbH, Z\"urich Research Laboratory, S\"aumerstrasse 4, 8803 R\"uschlikon, Switzerland}
\alsoaffiliation{Laboratory of Physical Chemistry,
ETH Z\"urich, Vladimir-Prelog-Weg 2, 8093 Z\"urich, Switzerland}
\author{Alberto Baiardi}
\affiliation{Laboratory of Physical Chemistry,
ETH Z\"urich, Vladimir-Prelog-Weg 2, 8093 Z\"urich, Switzerland}
\author{Markus Reiher}
\affiliation{Laboratory of Physical Chemistry,
ETH Z\"urich, Vladimir-Prelog-Weg 2, 8093 Z\"urich, Switzerland}
\email{markus.reiher@phys.chem.ethz.ch}
\author{Ivano Tavernelli}
\affiliation{IBM Research GmbH, Z\"urich Research Laboratory, S\"aumerstrasse 4, 8803 R\"uschlikon, Switzerland}
\date{February 2020}
\email{ita@zurich.ibm.com}

\newcommand\vertarrowbox[3][3ex]{
  \begin{array}[t]{@{}c@{}} #2 \\
  \left\uparrow\vcenter{\hrule height #1}\right.\kern-\nulldelimiterspace\\
  \makebox[0pt]{\scriptsize#3}
  \end{array}%
}

\begin{document}

% -- User-defined commands --
\newcommand{\quotes}[1]{``#1''}

\maketitle

\begin{abstract}
We introduce a framework for the calculation of ground and excited state energies of bosonic systems suitable for near-term quantum devices and apply it to molecular vibrational anharmonic Hamiltonians.
Our method supports generic reference modal bases and Hamiltonian representations, including the ones that are routinely used in classical vibrational structure calculations.
We test different parametrizations of the vibrational wave function, which can be encoded in quantum hardware, based either on heuristic circuits or on the bosonic Unitary Coupled Cluster \textit{Ansatz}.
In particular, we define a novel compact heuristic circuit and demonstrate that it provides the best compromise in terms of circuit depth, optimization costs, and accuracy.
We evaluate the requirements, number of qubits and circuit depth, for the calculation of vibrational energies on quantum hardware and compare them with state-of-the-art classical vibrational structure algorithms for molecules with up to seven atoms.
\end{abstract}

\section{Introduction}

Within the Born-Oppenheimer approximation, a molecular wave function is factorized as a product of an electronic part, which is the solution of the electronic Schr\"{o}dinger equation, and a vibro-rotational one, which is the solution of the nuclear Schr\"{o}dinger equation in the potential energy surface (PES) generated by sampling the eigenvalues of the electronic Schr\"{o}dinger equation for different geometries.

The nuclear Schr\"{o}dinger equation is usually solved in two steps, in analogy with its electronic counterpart. 
A single-particle basis (the basis functions are called, in this case, modals) is obtained either by the harmonic approximation applied to the PES or from a vibrational self-consistent field (VSCF)~\cite{bowman1978,carney1978,gerber1979,Gerber2013_Review} calculation.
Vibrational anharmonic correlations are added \textit{a-posteriori} with perturbative\cite{christiansen2003,Barone2005_VPT2} or variational approaches.
The latter include Vibrational Configuration Interaction (VCI)~\cite{bowman1979,thompson1980,christoffel1982,Rauhut2009_VCI-Fast} and Vibrational Coupled Cluster (VCC)~\cite{christiansen2004,Christiansen2009_Automatic-VCC} for highly-accurate anharmonic energies. 
Unlike perturbation theories, the accuracy of VCI and VCC can be systematically improved, but their applicability is limited to small molecules with up to about 10 atoms due to their unfavorable scaling with system size.
This unfavorable scaling can be tamed by pruning the VCI basis limiting, for instance, the maximum degree of excitation, or with precontraction algorithms.\cite{Rauhut2014_LiFClusters,Carrington2018_UracilNaphtalene,Bowman2019_EigenWater}
Such simplifications make calculations feasible for systems with up to 15-20 atoms.
Alternatively, the computational cost of VCI can be reduced with non-linear wave function parametrizations. 
This is the case, for example, of the vibrational Density Matrix Renormalization Group (vDMRG)~\cite{Baiardi2017_vDMRG,Baiardi2019_ExcitedStates-vDMRG} which encodes the wave function as a Matrix Product State~\cite{Reiher2020_Review}, or of VCC~\cite{christiansen2004}.

The emerging development of quantum computers has refreshed the prospect of computing energies of large molecules by leveraging the exponentially-large multi-qubit Hilbert space.
However, current quantum computers based on superconducting qubits technology have limited coherence times ($\approx100$~$\mu$s) and sizable gate error rates ($\approx2\times10^{-2}$ for a two-qubit gate), restricting the possible number of operations that can be executed to evolve a quantum state.
Under these limitations, hybrid quantum-classical algorithms are the most promising route to calculate molecular energies on quantum hardware.
In particular, the ground state energy of a general Hamiltonian can be obtained with quantum circuits of relatively low depth (i.e., a small number of operations) with the Variational Quantum Eigensolver (VQE)~\cite{peruzzo2014,yung2014,mcclean2016,wang2019}.
The VQE has already been applied in hardware calculation of the electronic ground state of small molecules~\cite{kandala2017,nam2019} and can also be extended to excited states~\cite{ollitrault2019,colless2018}.

Despite the extensive work on the applications of VQE to the solution of the electronic Schr\"{o}dinger equation, the extension to vibrational has not yet been fully investigated.
Molecular vibrations are described by Bose-Einstein statistics and, therefore, the modal basis must be mapped to the qubits by preserving such symmetry.
Moreover, any many-body expansion of a $L$-mode PES, contains, in principle, up to $L$-body coupling terms~\cite{Gerber1999_nMode,Kongsted2006_nModeRepresentation}. 
The potential energy operator of the nuclear Schr\"{o}dinger equation is therefore much more complex than the pairwise Coulomb interaction of the electronic Schr\"{o}dinger equation.
McArdle \textit{et al.}~\cite{mcardle2019} adapted the VQE to find the ground state of vibrational Hamiltonians of small molecules on a universal quantum computer based on the unitary extension of the VCC theory (UVCC).
In particular, they represent vibrational levels with the so-called compact mapping (the problem of mapping bosonic states to qubits has been discussed in details in a recent work~\cite{sawaya2019}).
Two key limitations hinder the application of the theory presented in Ref.~\citenum{mcardle2019} to complex vibrational Hamiltonians.
First, the algorithm can be applied only to vibrational ground states and, therefore, does not allow to access vibrational excitation energies that are key for vibrational spectroscopy. 
Second, it approximates the PES as power series of Cartesian-based normal modes, which relies on harmonic-oscillator eigenfunctions as basis functions.
This is an important limitation in the case of strongly anharmonic molecules, whose PES is represented by highly non-compact Taylor expansions.
For these systems, the VSCF modals will lead instead to more compact VCC and VCI expansions.

In the present paper, we design a general framework for the calculation of vibrational structures with a quantum algorithm.
We introduce the qubit encoding of the vibrational levels based on the generalized second quantization representation~\cite{christiansen2004,Wang2009_SQMCTDH} of the nuclear Schr\"{o}dinger equation that enables the potential to be expressed as a general $n$-mode expansion~\cite{Kongsted2006_nModeRepresentation}.
We then discuss the parametrization of the vibrational wave function introducing a new quantum circuit \textit{Ansatz} as a compact approximation of UVCC.
We emphasize that, although UCC has mostly been applied in electronic-structure calculations~\cite{Bartlett2006_UCC-Review,bartlett2007,Lee2019_Generalized-UCC,Evangelista2019_UCC} and its extension to vibrational quantum computation has hardly been explored~\cite{mcardle2019}, its use for the solution of the vibronic problem is very promising.
We discuss how vibrational excited states can be targeted with equation-of-motion (EOM)-based UVCC algorithms~\cite{ollitrault2019}.
Finally, we also discuss the scaling of the UVCC resources in terms of qubits and gate counts as a function of the molecular size and show the resources necessary to compute the vibrational structure of molecules with up to five atoms.
The proposed framework offers us the possibility to estimate the hardware requirements that will allow to reach quantum advantage over classical vibrational-structure calculations using near-term quantum computers.

\section{Theory}

\subsection{Second quantization theories for molecular vibrations}

The real-space representation of the Watson Hamiltonian for the $L$ modes of a molecular sytem can be written as
\begin{equation}
  \mathcal{H}_\text{vib}(Q_1, \ldots, Q_L) = - \frac{1}{2} \sum_{l=1}^{L} \frac{\partial^2}{\partial Q_l^2}
    + V(Q_1, \ldots, Q_L)
  \label{eq:VibHam_Coordinates}
\end{equation}
where $Q_l$ are the harmonic mass-weighted normal coordinates and the Coriolis couplings~\cite{WilsonBook,PapousekBook} have been neglected.
$\mathcal{H}_\text{vib}$ must be mapped to an operator that acts on the states of a given set of $N_q$ qubits in order to calculate its eigenfunctions on quantum hardware.
In electronic structure calculations, the mapping is achieved by expressing the non-relativistic electronic Hamiltonian in second quantization, \textit{i.e.} by projecting it onto the complete set of antisymmetrized occupation number vectors (ONV) generated by a given (finite) set of orbitals.
To encode the vibrational Hamiltonian of Eq.~(\ref{eq:VibHam_Coordinates}) in terms of the second quantization operators defined in Eq.~(\ref{eq:SQBosonicDefinition}), we expand the potential $V(Q_1, \ldots, Q_L)$ with the $n$-body expansion~\cite{Gerber1999_nMode,Kongsted2006_nModeRepresentation}, as follows:
\begin{equation}
  V(Q_1, \ldots, Q_L) = V_0 + \sum_{l=1}^L V^{[l]}(Q_l) + \sum_{l<m}^L V^{[l,m]}(Q_l, Q_m)
    + \sum_{l<m<n}^L V^{[l,m,n]}(Q_l, Q_m, Q_n) + \ldots
  \label{eq:ManyBodyPotential}
\end{equation}
where $V_0$ is the electronic energy of the reference geometry, the one-mode term $V^{[l]}(Q_l)$ represents the variation of the PES upon change of the $l$-th normal coordinate from the equilibrium position.
Similarly, the two-body potential $V^{[l,m]}(Q_l, Q_m)$ represents the change in the exact PES upon a simultaneous displacement along the $l$-th and $m$-th coordinates~\cite{Kongsted2006_nModeRepresentation}. 
The exact representation of a PES for an $L$-mode system requires an $L$-body expansion.
Often, including terms up to three-body in the $L$-body expansion is sufficient to obtain an accuracy of about 1~cm$^{-1}$.

A representation of Eq.~(\ref{eq:VibHam_Coordinates}) that is suitable to encode on a quantum computer can be obtained with the so-called canonical quantization~\cite{WilsonBook} that maps the $l$-th normal coordinate $Q_l$ and its conjugate momentum $P_l$ to a pair of bosonic creation and annihilation operators ($a_l^+$ and $a_l$) defined as

\begin{equation}
 \begin{aligned}
   Q_l &= \frac{1}{\sqrt{2}} \left( a_l^+ + a_l \right) \\
   P_l &= \frac{i}{\sqrt{2}} \left( a_l^+ - a_l \right) \, ,
 \end{aligned}
 \label{eq:CanonicalQuantization}
\end{equation}
where the $a_l^+$/$a_l$ operators are defined as

\begin{equation}
 \begin{aligned}
   a_l^+ \ket{ n_1 \cdots n_l \cdots n_L } &= \sqrt{n_l+1} \; \ket{ n_1 \cdots n_l+1 \cdots n_L} \\
   a_l   \ket{ n_1 \cdots n_l \cdots n_L } &= \sqrt{n_l}   \; \ket{ n_1 \cdots n_l-1 \cdots n_L }.
 \end{aligned}
 \label{eq:CanonicalSQOperators}
\end{equation}

Each index of the ONV $\ket{ n_1 \cdots n_L }$ is associated to a mode and $n_l$ is the degree of excitation of the $l$-th mode.
Different vibrational-structure methods have been derived based on this canonical representation,\cite{Hirata2014_StochasticVSCF,Hirata2014_NormalOrdered} including VCC,\cite{Prasad1994_Original,Prasad2008_VCC-Response,Hirata2018_SimilarityTransformedVCC,Hirata2018_HighOrderVCC}  although it is not flexible enough to target strongly anharmonic systems. 
In this formalism the PES $V(Q_1,\ldots,Q_L)$ is expressed as a power series to encode it in a second quantization format based on Eq.~(\ref{eq:CanonicalQuantization}).
In addition, the operators of Eq.~(\ref{eq:CanonicalSQOperators}) imply that the reference basis set for every mode $l$ are the harmonic oscillator eigenfunctions.
However, such a basis does not lead to a compact representation of vibrational wave functions for strongly anharmonic systems for which modals obtained, for instance, from VSCF \cite{bowman1978,bowman1986} are better suited.

A more flexible second quantization form is the so-called $n$-mode representation introduced by Christiansen~\cite{Christiansen2004_nMode}.
Instead of labelling each basis function with a single integer, as in Eq.~(\ref{eq:CanonicalSQOperators}), we expand each mode $l$ into a basis of $N_l$ modals (labelled as $i_1 \cdots i_{N_l}$) which generates an ONV basis for that mode.
Let us consider the following, general VCI expansion

\begin{equation}
  \ket{\Psi} = \sum_{k_1=1}^{N_1} \cdots \sum_{k_L=1}^{N_L} C_{k_1,\ldots,k_L} 
    \phi_{k_1}^{(1)}(Q_1) \cdots \phi_{k_L}^{(L)}(Q_L) \, ,
  \label{eq:VCI_parametrization}
\end{equation}
where each mode $l$ is described by the $N_l$-dimensional basis set $S_l$ defined as
\begin{equation}
  S_l = \{ \phi_1^{(l)} (Q_l) , \ldots , \phi_{N_l}^{(l)} (Q_l) \} \, .
  \label{eq:VibrationalBasis}
\end{equation}
The many-body basis function $\phi_{k_1}^{(1)}(Q_1) \cdots \phi_{k_L}^{(L)}(Q_L)$ can be encoded as an ONV as
\begin{equation}
  \phi_{k_1}(Q_1) \cdots \phi_{k_L}(Q_L)
                      \equiv  \ket{0_1 \cdots 1_{k_1} \cdots 0_{N_1},
                                   0_1 \cdots 1_{k_2} \cdots 0_{N_2}, 
                                   \cdots , 
                                   0_1 \cdots 1_{k_L} \cdots 0_{N_L}} \, .
  \label{eq:ONV_Christiansen}
\end{equation}
The full ONV is then given by the expression in Eq.~(\ref{eq:ONV_Christiansen}) where different ONV subspaces are separated by a comma.
For each ONV space, we sort the modals in decreasing order of energy.
Each mode is described by one and only one basis function, therefore the occupation of each ONV subspace is one.

Based on the representation given in Eq.~(\ref{eq:ONV_Christiansen}), we introduce a pair of creation and annihilation operators per mode $l$ \textit{and} per basis function $k_l$ defined as:

\begin{equation}
  \begin{aligned}
    a_{k_l}^\dagger \ket{ \cdots, 0_1 \cdots 0_{k_l} \cdots 0_{N_l}, \cdots} 
      &=  \ket{ \cdots, 0_1 \cdots 1_{k_l} \cdots 0_{N_l}, \cdots} \\
    a_{k_l}^\dagger \ket{ \cdots, 0_1 \cdots 1_{k_l} \cdots 0_{N_l}, \cdots} &=  0 \\
    a_{k_l} \ket{ \cdots, 0_1 \cdots 1_{k_l} \cdots 0_{N_l}, \cdots} 
     &= \ket{ \cdots, 0_1 \cdots 0_{k_l} \cdots 0_{N_l}, \cdots} \\
    a_{k_l} \ket{ \cdots, 0_1 \cdots 0_{k_l} \cdots 0_{N_l}, \cdots} &=  0 \\
  \end{aligned}
  \label{eq:SQBosonicDefinition}
\end{equation}
with
\begin{equation}
  \begin{aligned}
    \left[ a_{k_l}^\dagger, a_{h_m}^\dagger \right] &= 0 \\
    \left[ a_{k_l}, a_{h_m} \right] &= 0 \\
    \left[ a_{k_l}^\dagger, a_{h_m} \right] &= \delta_{l,m} \, , \delta_{k_l,h_m}
  \end{aligned}
  \label{eq:BosonicCommutationRule}
\end{equation}
This formalism was introduced for VCC~\cite{christiansen2004} and later applied to the multi-configurational time-dependent Hartree method \cite{Wang2009_SQMCTDH}.
The second quantization form of Eq.~(\ref{eq:VibHam_Coordinates}) obtained by expressing the potential as in Eq.~(\ref{eq:ManyBodyPotential}) reads\cite{Christiansen2004_nMode}

\begin{equation}
 \begin{aligned}
  \mathcal{H}_\text{vib}^{SQ} =& \sum_{l=1}^L 
    \sum_{k_l,h_l}^{N_l} \langle \phi_{k_l} | T(Q_l) + V^{[l]}(Q_l) | \phi_{h_l} \rangle a_{k_l}^+ a_{h_l} \\
 +& \sum_{l<m}^L \sum_{k_l,h_l}^{N_l} \sum_{k_m,h_m}^{N_m}
    \langle \phi_{k_l} \phi_{k_m} | V^{[l,m]}(Q_l, Q_m) | \phi_{h_l} \phi_{h_m} \rangle 
    a_{k_l}^+ a_{k_m}^+ a_{h_l} a_{h_m} + \cdots
 \end{aligned}
 \label{eq:SQHam_NMode}
\end{equation}

Unlike its electronic-structure counterpart, Eq.~(\ref{eq:SQHam_NMode}) contains in general coupling terms higher than two-body. Therefore, the number of Pauli terms to be evaluated on the quantum computer scales as $\mathcal{O}(N^{2n})$ for a $n$-body truncation, where $N$ is the overall number of modals.
We highlight that any PES can be encoded in the $n$-mode second quantization format provided that the integrals of the PES over the modals are available.
Eq.~(\ref{eq:SQHam_NMode}) is therefore not restricted to PESs expressed as a power series.

\subsection{Wave function parametrization}

The second quantization formalism introduced in the previous section allows one to express the VCI expansion in terms of ONVs constructed with modals that do not rely on the harmonic approximation.
The encoding of such ONVs on a quantum computer is straightforward if based on a one-to-one correspondence between modals and qubits. 
This mapping extends the \quotes{direct mapping} of Ref.~\citenum{sawaya2019} beyond harmonic reference basis sets.
The $N_l$ modals for a given mode $l$ are represented by a $N_l$-qubit register. 
We sort the modals in decreasing order of energy.
Therefore, the lowest-energy configuration is represented by the ONV $\ket{0_1 \cdots 1_{N_1}, 0_1 \cdots 1_{N_2}, \cdots, 0_1 \cdots 1_{N_L}}$ and is obtained by applying an $X$ gate on the first qubit of each mode register initialized in the vacuum state.
The correlated wave function is obtained from the reference state by applying a set of excitation operators defined by a given wave function \textit{Ansatz}.

In VQE-based electronic-structure quantum-computing two main strategies are available to prepare the wave function.
The first is based on the CC method and, more precisely, on its unitary formulation (UCC). 
It provides an intuitive expansion of the wave function in terms of excitation operators controlled by an efficiently parametrized circuit.~\cite{peruzzo2014,mcardle2018,barkoutsos2018,lee2018}
However, this circuit comprises a large number of 2-qubit gates (CNOT gates) and, hence, its practical use is limited by coherence time and the gate error rates.
The second approach does not have a classical equivalent and is tailored to quantum hardware.
A heuristic wave function \textit{Ansatz} is built concatenating parametrized single-qubit rotations and entangling blocks.~\cite{barkoutsos2018,kandala2017}
The number of parameters can be increased by repeating the same set of operations (but with independent parameters) $d$ times (where $d$ refers to the circuit depth) to reach the desired accuracy for the ground state energy.

The same strategies can be followed for preparing vibrational wave functions. 
The UVCC circuit can be obtained from the unitary version of the VCC~\cite{christiansen2004,mcardle2019} \textit{Ansatz}:
\begin{equation}
  \ket{\Psi} = e^{\mathcal{T}-\mathcal{T}^{\dagger}} \ket{\Psi_{\text{ref}}} \, ,
  \label{eq:UVCC_Parametrization}
\end{equation}
where $\ket{\Psi_{\text{ref}}}$ is the reference ONV. $\mathcal{T}$ is the cluster operator (and $\mathcal{T}^\dagger$ its adjoint) expressed here up to second order as
\begin{equation}
  \mathcal{T} = \mathcal{T}_1 + \mathcal{T}_2  \, ,
  \label{eq:ClusterOperator}
\end{equation}
with
\begin{align}
   \mathcal{T}_1 &= \sum_l^L \sum_{h_l,k_l}^{N_l} 
      \theta_{h_l,k_l} a^{\dagger}_{h_l} a_{k_l} 
   \label{eq:ClusterOperator_ExcitationClass_T1} \\
   \mathcal{T}_2 &= \sum_{l<m}^L \sum_{h_l,k_l}^{N_l} \sum_{h_m,k_m}^{N_m} 
      \theta_{h_l k_l h_m k_m} a^{\dagger}_{h_l} a^{\dagger}_{h_m} a_{k_m} a_{k_l}
   \label{eq:ClusterOperator_ExcitationClass_T2} 
\end{align}
and $(h_l,k_l)$ and $(h_m,k_m)$ label couple of modals for the modes $l$ and $m$, respectively.
The bosonic $a^{\dagger}_{k_l}$ and $a_{k_l}$ operators (see Eq.~(\ref{eq:SQBosonicDefinition})) are mapped to the Pauli operators $\sigma^+_{k_l}=\sigma^x_{k_l} + i \sigma^y_{k_l}$ and $\sigma^-_{k_l}=\sigma^x_{k_l} - i \sigma^y_{k_l}$, respectively.
In this way, the exponential operator of Eq.~(\ref{eq:UVCC_Parametrization}) can be factorized with a Trotter expansion and expressed as a product of quantum gates.
Compared to the fermion-to-qubit mappings used in electronic-structure calculations (where the antisymmetry of the wave function is encoded in the circuit by applying for instance the Jordan Wigner transformation~\cite{jordan1993}) the circuit depth for the implementation of the UVCC \textit{Ansatz} is greatly reduced.

Among the heuristic circuits designed for electronic structure calculation~\cite{barkoutsos2018,kandala2017}, we consider here the SwapRZ \textit{Ansatz} defined as
\begin{equation}
  \ket{\Psi'} = e^{i\mathcal{T}'} \ket{\Psi_{\text{ref}}} \, ,
  \label{eq:SwapRZ}
\end{equation}
with
\begin{equation}
  \mathcal{T}' = \sum_{i<j}^{N_q} \theta_{i,j} (X_iX_j + Y_iY_j) \, ,
  \label{eq:SwapRZ_SingleGate}
\end{equation}
where we use the notation $X_i$ and $Y_i$ for the $\sigma_x$ and $\sigma_y$ Pauli matrices acting on qubit $i$ and $N_q$ is the overall number of qubits.
In addition, two layers of single-qubit RZ rotations parametrized by an extra set of $2N_q$ angles are applied before and after the entangler block required which encodes the expansion in Eq.~(\ref{eq:SwapRZ}).
The SwapRZ circuit ensures that the expansion for $\ket{\Psi'}$ in made of ONVs with $L$ and only $L$ occupied modals. However, since the circuit also entangles pair of qubits describing modals belonging to different modes, this procedure does not ensure that a single modal per mode will be occupied.
This is not the case for the UVCCS Ansatz (UVCC with $\mathcal{T} = \mathcal{T}_1$), where single excitations are confined to the modal space of the same mode (see Eq.~(\ref{eq:ClusterOperator_ExcitationClass_T1})).
Simultaneous excitations of two different modes (as those included in $\mathcal{T}_2$) are not explicitly captured by the SwapRZ \textit{Ansatz} with depth 1 and, therefore, we expect that deeper circuits are required to accurately represent the wave function.

Another strategy proposed in the context of electronic structure~\cite{barkoutsos2018} is to build the circuit from a layer of RY and RZ rotations on each qubit followed by a block of CNOT gates entangling all qubits. 
We refer to the resulting circuit as RYRZ.
Note that rotations around the Y axis of each qubit induce a change in both the overall modals occupation and the individual occupation of each mode.

To constrain the optimization to the correct symmetry subspace, both SwapRZ and RYRZ heuristic circuits must be combined to a modified Hamiltonian $\mathcal{H}_\text{vib}'$ where a penalty function is added to increase the energy of the states with unphysical occupation\cite{barkoutsos2018},
\begin{equation}
  \mathcal{H}_\text{vib}' = \mathcal{H}_\text{vib} 
    + \mu \sum_{l=1}^L \left( \bra{\Psi} \mathcal{N}_l \ket{\Psi} - 1\right)^2 \, ,
  \label{eq:ModifiedHamiltonian}
\end{equation}
where $\mu$ is an arbitrary parameter and the number operator $\mathcal{N}_l$ for mode $l$ is defined as
\begin{equation}
  \mathcal{N}_l = \sum_{k_l=1}^{N_l} a_{k_l}^\dagger a_{k_l} \, .
  \label{eq:NumberOperator-Definition}
\end{equation}
In the next section we will show that the optimization of the wave function is much more efficient with the UVCC Ansatz than with the heuristic ones.
Therefore, it is desirable to derive a quantum circuit inspired by UVCC that involves a smaller number of CNOT gates, and hence is more suited for near-term quantum calculations.

Given the current coherence time and gate error rates, it is challenging to include double excitations within the UVCC circuit.
In fact, each element of $\mathcal{T}_2$ can be decomposed as:
\begin{equation}
 \begin{aligned}
  \sigma^{+}_i\sigma^{+}_j\sigma^{-}_k\sigma^{-}_l - \text{ c.c.} 
    =  2i(& X_iY_jX_kX_l + Y_iX_jX_kX_l + Y_iY_jX_kY_l + Y_iY_jY_kX_l \\
        - & X_iX_jX_kY_l - X_iX_jY_kX_l - X_iY_jY_kY_l - Y_iX_jY_kY_l) \, .
 \end{aligned}
 \label{eq:EntanglingBlock}
\end{equation}
Therefore, the corresponding quantum circuit obtained after exponentiation of the operator given in Eq.~(\ref{eq:EntanglingBlock}) and its Trotterization contains $8\times 6$ CNOT gates per excitation.

\begin{figure}[htbp!]
  \centering
  \includegraphics[width = \textwidth]{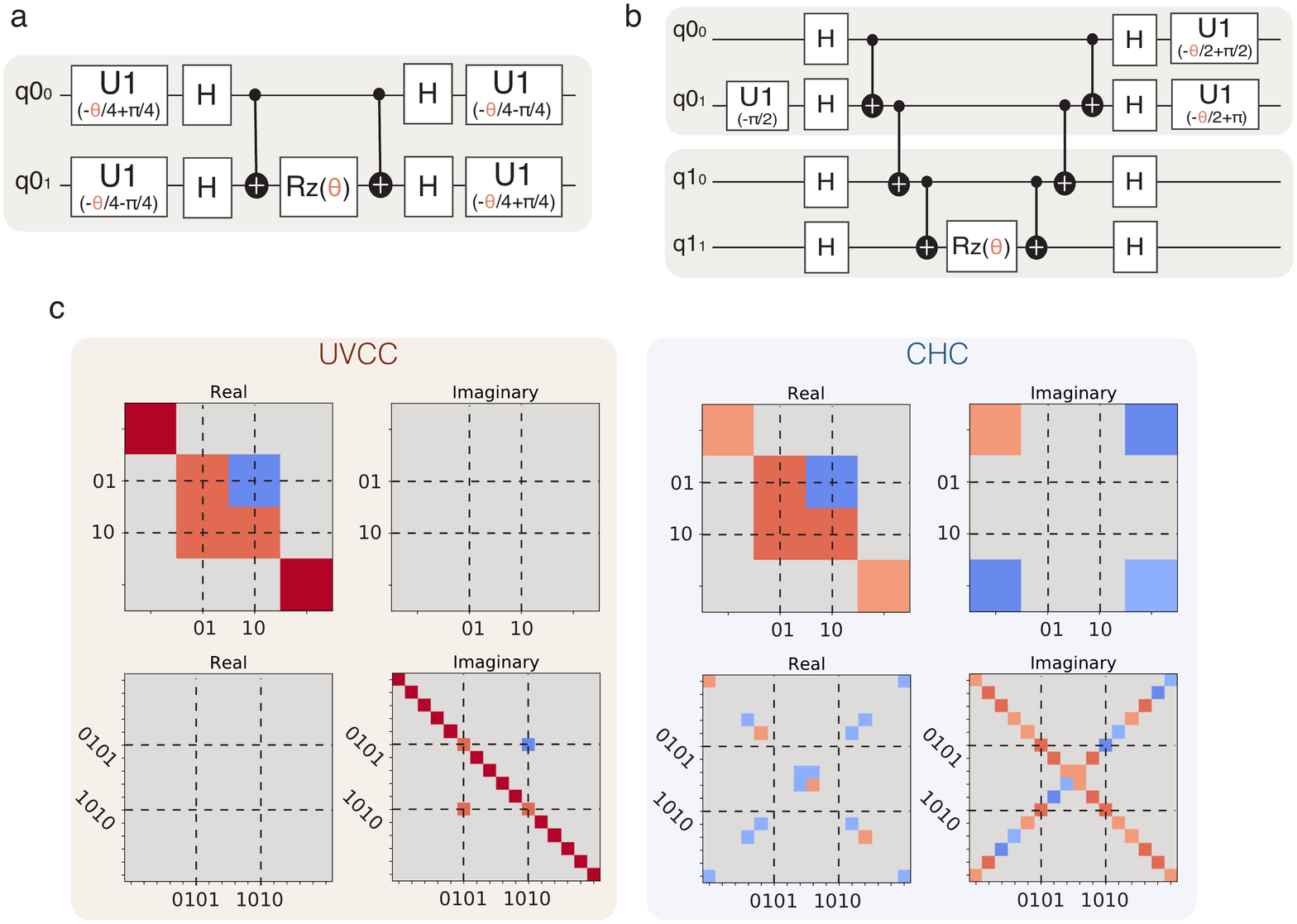}
  \caption{\textbf{a.} CHC circuit approximating a $\mathcal{T}_1-\mathcal{T}_1^{\dagger}$ type of excitation; \textbf{b.} CHC circuit approximating a $\mathcal{T}_2-\mathcal{T}_2^{\dagger}$ type of excitation; \textbf{c} Unitary matrices corresponding to a single (top) and a double (bottom) excitation with UVCC and the corresponding approximation with CHC with RZ rotation angle $\theta = \pi/4$. The matrix elements are represented with a color according to the map $\in \{$-1 (blue) to 1 \text{(red)}$\}$.} 
  \label{fig:uvcc_chc_circ_&_mat}
\end{figure}

Combining the considerations above, we propose to approximate the UVCCSD circuit with a more compact heuristic \textit{Ansatz} that we name Compact Heuristic for Chemistry (CHC). This circuit exploits the fact that the relations

\begin{equation}
  |\bra{a_i^+ a_m \Psi_{\text{ref}}} e^{i\theta_m^i \, X_iX_m}\ket{\Psi_{\text{ref}}}|^2 
    = | \bra{a_i^+ a_m \Psi_{\text{ref}}} e^{ \theta_m^i (\sigma^{+}_i\sigma^{-}_m -
        \text{c.c.})}\ket{\Psi_{\text{ref}}} |^2
  \label{eq:CHC_Singles}
\end{equation}
and
\begin{equation}
  | \bra{a_i^+ a_j^+ a_m a_n \Psi_{\text{ref}}} e^{i\theta_{m,n}^{i,j} \, X_iX_jX_mX_n} \ket{\Psi_{\text{ref}}}|^2
    = | \bra{a_i^+ a_j^+ a_m a_n \Psi_{\text{ref}}}
        e^{\theta_{m,n}^{i,j} (\sigma^{+}_i\sigma^{+}_j\sigma^{-}_m\sigma^{-}_n 
        - \text{c.c.})}\ket{\Psi_{\text{ref}}}|^2
  \label{eq:CHC_Doubles}
\end{equation}
hold for indices $m$, $n$ and $i$, $j$ corresponding to occupied and unoccupied modals in the reference state, respectively.
However the relative phase of the configurations in the resulting state differs. 
To correct this phase difference, we introduce the compact circuits $U_s^{m,i}(\theta_m^i)$ and $U_d^{m,n,i,j}(\theta_{m,n}^{i,j})$, presented in Figures~\ref{fig:uvcc_chc_circ_&_mat}a and \ref{fig:uvcc_chc_circ_&_mat}b respectively, for which the above relations become
\begin{equation}
  U^{m,i}_s(\theta_m^i) \ket{\Psi_{\text{ref}}} 
    = e^{\theta_m^i (\sigma^{+}_i\sigma^{-}_m - \text{c.c.})}\ket{\Psi_{\text{ref}}} \, ,
  \label{eq:CHC_1}
\end{equation}
and
\begin{equation}
  U^{m,n,i,j}_d(\theta_{m,n}^{i,j}) \ket{\Psi_{\text{ref}}} 
    = e^{\theta_{m,n}^{i,j} 
      (\sigma^{+}_i\sigma^{+}_j\sigma^{-}_m\sigma^{-}_n - \text{c.c.})}\ket{\Psi_{\text{ref}}}
  \label{eq:CHC_2}
\end{equation}
for variable parameters $\theta_m^i$ and $\theta_{m,n}^{i,j}$.
The shortcomings of CHC are related to the fact that the $U^{m,i}_s(\theta_m^i)$ and $U^{m,n,i,j}_d(\theta_{m,n}^{i,j})$ operators are applied sequentially for each excitation. Therefore, in general, the circuit that corresponds to a given excitation (shown in Fig~\ref{fig:uvcc_chc_circ_&_mat}) is not directly applied on the reference state $\ket{\Psi_\text{ref}}$ but rather on a superposition state generated through the previous excitations.
This can produce unphysical configurations with the wrong number of particles for each mode.
For weakly correlated systems, we expect CHC to act very similarly to UVCC as the weight of the unphysical configurations will be negligible.
However, this approximation deteriorates when working with strongly correlated systems, \textit{i.e.} with a VCI wave function that is not dominated by a single configuration.
The decisive advantage of CHC is that the number of CNOT gates is reduced by approximately one order of magnitude compared to UCC, allowing therefore the computation of larger systems.
We note that the unphysical configurations can also be efficiently projected out with a Monte-Carlo inspired scheme described in Ref.~\citenum{mazzola2019}. 
The scaling and performance of the CHC circuit are presented in Section~\ref{sec:results_gs}.

\subsection{Extension to excited states}

The calculation of the ground state is not sufficient for most vibrational-structure calculations for which vibrational excitation energies must also be considered.
This is the case for the simulation of vibrational spectra, in which peaks are located at transition frequencies.
In quantum computing, the calculation of excited states can be performed with the quantum EOM (qEOM) algorithm\cite{ollitrault2019} in which a vibrational excited state $\ket{n}$ is expressed as $\ket{n} = \text{O}_{n}^{\dagger} \ket{0}$, where $\ket{0}$ indicates the vibrational ground state and $\text{O}_{n}^{\dagger}$ is defined as

\begin{equation}
    \text{O}_{n}^{\dagger} 
      = \sum_{\alpha} \sum_{\mu_{\alpha}} \left[ X_{\mu_{\alpha}}^{(\alpha)} (n) \mathcal{E}_{\mu_{\alpha}}^{(\alpha)} 
          - Y_{\mu_{\alpha}}^{(\alpha)}(n) (\mathcal{E}_{\mu_{\alpha}}^{(\alpha)})^{\dagger} \right] \, .
    \label{eq:O_expansion}
\end{equation}

In Eq.~(\ref{eq:O_expansion}), $\alpha$ is the order of the excitation and the collective index $\mu_{\alpha}$ runs over all modals involved in the excitation. The excitation operators $\mathcal{E}_{\mu_{\alpha}}^{(\alpha)}$ are sequences of creation and annihilation operators. For instance $\mathcal{E}_{\mu_2}^{(2)} = a_{k_l}^{\dagger} a_{h_m}^{\dagger} a_{0_m} a_{0_l} $, with $l<m$ running over the number of modes and $k_l,h_m$ over the number of modals for modes $l$ and $m$, respectively. In addition, $X_{\mu_{\alpha}}^{(\alpha)}$ is the expansion coefficient for the $\mathcal{E}_{\mu_{\alpha}}^{(\alpha)}$ excitation operator, and $Y_{\mu_{\alpha}}^{(\alpha)}$ is the coefficient for the corresponding de-excitation.
The excitation energy $E_{0n}$ is found by solving classically the following pseudo-eigenvalue problem

\begin{equation} 
  \begin{pmatrix}
    \text{\textbf{M}} & \text{\textbf{Q}}\\ 
    \text{\textbf{Q*}} & \text{\textbf{M*}}
  \end{pmatrix}
  \begin{pmatrix}
    \text{\textbf{X}}_n\\ 
    \text{\textbf{Y}}_n
  \end{pmatrix}
  = E_{0n}
  \begin{pmatrix}
    \text{\textbf{V}} & \text{\textbf{W}}\\ 
    -\text{\textbf{W*}} & -\text{\textbf{V*}}
  \end{pmatrix}
  \begin{pmatrix}
    \text{\textbf{X}}_n\\ 
    \text{\textbf{Y}}_n
  \end{pmatrix} ,
  \label{eq:XnYn}
\end{equation}
where the matrix elements

\begin{equation}
 \begin{aligned}
  &M_{\mu_{\alpha}\nu_{\beta}} = \bra{0} [(\mathcal{E}_{\mu_{\alpha}}^{(\alpha)})^{\dagger},\mathcal{H}, \mathcal{E}_{\nu_{\beta}}^{(\beta)}]\ket{0}, \\
  & Q_{\mu_{\alpha}\nu_{\beta}} = -\bra{0} [(\mathcal{E}_{\mu_{\alpha}}^{(\alpha)})^{\dagger}, \mathcal{H}, (\mathcal{E}_{\nu_{\beta}}^{(\beta)})^{\dagger}]\ket{0},\\
  & V_{\mu_{\alpha}\nu_{\beta}} = \bra{0} [(\mathcal{E}_{\mu_{\alpha}}^{(\alpha)})^{\dagger}, \mathcal{E}_{\nu_{\beta}}^{(\beta)}]\ket{0},\\
  &W_{\mu_{\alpha}\nu_{\beta}} = -\bra{0} [(\mathcal{E}_{\mu_\alpha}^{(\alpha)})^{\dagger}, (\mathcal{E}_{\nu_{\beta}}^{(\beta)})^{\dagger}]\ket{0}.
 \end{aligned}
 \label{eq:qEOM_MatrixElements}
\end{equation}
are evaluated with the approximated ground state wavefunction generated with the quantum circuit.
The accuracy of qEOM can be improved systematically by increasing the maximum excitation order $\alpha$.
The inclusion of excitations with $\alpha > 2$ may become important for PESs that require the inclusion of three- and higher-order terms in the $n$-body expansion of Eq.~(\ref{eq:ManyBodyPotential}).
This increases the size of the pseudo-eigenvalue problem and, therefore, the number of measurements to be performed on the quantum hardware.

\section{Results}

As an example system, we choose the carbon dioxide molecule, which is well studied with traditional approaches\cite{Hirata2007_CO2-PES,Hirata2013_CO2-Pressure,Bowman2017_CO2-Copper}.
We also estimate the quantum computing resources needed for the simulation of two larger molecules, namely formaldehyde (H$_2$CO) and formic acid (HCOOC).

\subsection{Ground state calculations with state-of-the-art approaches}
\label{sec:results_gs}

We study the UVCC, SwapRZ and RYRZ wave function approaches on the PES of $\rm{CO}_2$ defined by the bending and the symmetric stretching modes.
We describe the system Hamiltonian in second quantization as in Eq.~(\ref{eq:SQHam_NMode}) with two modals per mode where the reference modal basis (see Eq.~(\ref{eq:VibrationalBasis})) is obtained as eigenfunctions of the one-body Hamiltonian

\begin{equation}
  \left( T(Q_l) + V^{[l]}(Q_l) \right) \phi_{k_l}(Q_l)
    = \epsilon_{k_l} \phi_{k_l}(Q_l) \qquad (l=1,2) \, .
  \label{eq:PseudoSCF}
\end{equation}

One could include vibrational correlations in Eq.~(\ref{eq:PseudoSCF}) by introducing the averaged two- and higher-order couplings with the other modes in the potential operator, as done in VSCF~\cite{bowman1986}.
The resulting set of modals would lead to more compact VCI and VCC expansions than those based on the modals from Eq.~(\ref{eq:PseudoSCF}).
However, modals obtained from Eq.~(\ref{eq:PseudoSCF}) are sufficient to assess the quality of our algorithm and demonstrate that bases different from the harmonic ones can be used.
We optimize the ground-state equilibrium geometry of $\rm{CO}_2$ using density functional theory with the B3LYP exchange-correlation functional~\cite{becke1993} and the cc-pVTZ~\cite{Dunning1989_Basis} basis set.
We approximate the PES with a quartic force field and calculate the anharmonic force fields by semi-numerical differentiation of the analytical Hessian as implemented in \texttt{Gaussian}~\cite{g09}.

\begin{figure}[htbp!]
  \centering
  \includegraphics[width = 0.8\textwidth]{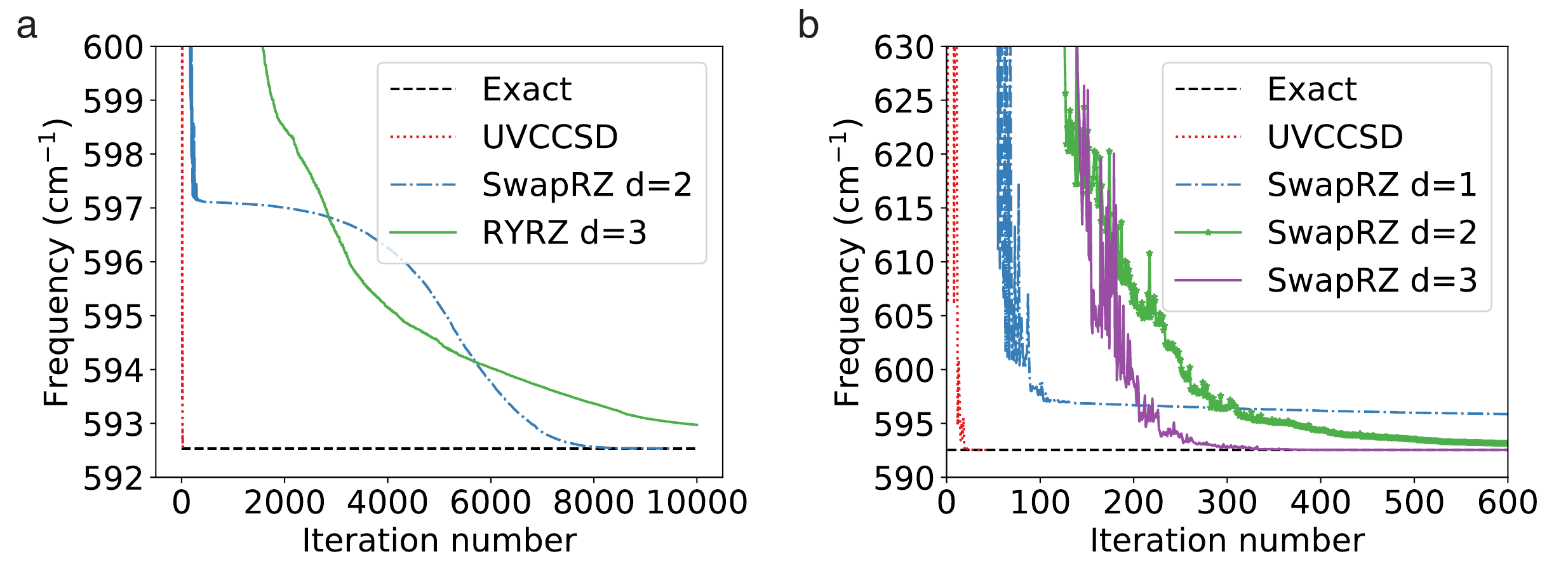}
  \caption{\textbf{a.} Convergence of the VQE algorithm for the calculation of the vibrational ground state of CO$_2$ in the 2 modes and 2 modals per node representation of the PES. A penalty term (Eq.~(\ref{eq:ModifiedHamiltonian})) is added to the Hamiltonian in order to enforce the conservation of one modal per mode. \textbf{b.} Same as in \textbf{a.} but without the addition of the penalty term.}
  \label{fig:known_circuits}
\end{figure}

All VQE calculations are run with \texttt{Qiskit}~\cite{Qiskit}. For the SwapRZ and RYRZ circuits, we set the penalty term $\mu$ in Eq.~(\ref{eq:ModifiedHamiltonian}) to $10^5$ (we find empirically that values below this threshold do not proscribe the convergence to the vacuum state).
For all VQE simulations, we apply the exact unitary matrix representation of the circuit on the reference ONV without taking into account sampling, decoherence and gate noise.
The results of the simulations are reported in Figure~\ref{fig:known_circuits}a.

We note that the convergence for both the SwapRZ and the RYRZ circuits is generally slower than with UVCCSD and convergence is reached only after 8000 iterations for SwapRZ, and after more than 10000 iterations for RYRZ. 
We recall that UVCC is the only occupation-conserving \textit{Ansatz}, therefore this pronounced difference can be due to the modified Hamiltonian of Eq.~(\ref{eq:ModifiedHamiltonian}).
The RYRZ circuit cannot be used without the penalty term since it does not conserve the overall modal occupation, and therefore it would converge to the vacuum state.
Conversely, even if the SwapRZ does not ensure that only one modal per mode is occupied, the energy gap between two modals is high enough to prevent the algorithm to converge to states with the incorrect occupation such as, for instance, the state $\ket{11,00}$.
Hence, we combine the SwapRZ \textit{Ansatz} with the original non-modified Hamiltonian to assess the effect of the addition of the penalty term, and compare the results with the UVCC approach. 
We run a VQE with the SwapRZ circuit for depths 1, 2 and 3. The number of entangling gates is 56, 24, 48, and 72 for UVCC, SwapRZ1, SwapRZ2, and SwapRZ3, which corresponds to a number of variational parameters of 3, 14, 24, and 34, respectively.
Fig.~\ref{fig:known_circuits}b shows that, even if the optimization is faster with the unmodified Hamiltonian, UVCC still outperforms SwapRZ for all depths.
This suggests that the explicit inclusion of double-excitations in the wave function \textit{Ansatz} is crucial in order to obtain reliable energies, even if this also leads to deeper circuits due to the presence of multiple CNOT gates.

\subsection{Performance of the CHC \textit{Ansatz}}

To reduce the circuit depth without sacrificing the accuracy, we approximate the UVCC wave function with the CHC \textit{Ansatz} (Eqs.~(\ref{eq:CHC_1}) and (\ref{eq:CHC_2})).
The scaling of UVCC and CHC in terms of number of CNOT gates and number of parameters is shown on Table~\ref{tab:ressources_circuit} for three molecules: CO$_2$, H$_2$CO and HCOOC.
For all these cases, we constructed the reference modal basis as described for CO$_2$, \textit{i.e.} from the eigenfunctions of the one-body Hamiltonian of Eq.~(\ref{eq:PseudoSCF}), based on geometries optimized with B3LYP/cc-pVTZ and calculating the anharmonic quartic force field by including all third-order and the semi-diagonal fourth-order terms.
\begin{table}[]
  \centering
  \begin{tabular}{c c c c c c}
    \hline
    \hline
    Molecule & Modes & Modals & CX UVCC & CX CHC & Parameters\\
    \hline
    $\rm{CO}_2$ & 4 & \makecell{2\\4\\6\\8\\10} & \makecell{304\\2640\\7280\\14224\\23472} & \makecell{44\\348\\940\\1820\\2988} & \makecell{10\\66\\170\\322\\522} \\
    \hline
    $\rm{H}_2\rm{CO}$ & 6 &  \makecell{2\\4\\6\\8\\10} & \makecell{744\\6552\\18120\\35448\\58536} & \makecell{102\\846\\2310\\4494\\7398} & \makecell{21\\153\\405\\777\\1269} \\
    \hline
    $\rm{HCOOH}$& 9 &  \makecell{2\\4\\6\\8\\10} & \makecell{1764\\15660\\43380\\84924\\140292} & \makecell{234\\1998\\5490\\10710\\17658} & \makecell{45\\351\\945\\1827\\2997} \\
    \hline\hline
  \end{tabular}
  \caption{Quantum circuit resource estimation for the calculation of the ground-state vibrational energy of $\rm{CO}_2$, $\rm{H}_2\rm{CO}$, and $\rm{HCOOH}$ with the UVCC and CHC approaches including single and double excitations. The number of CNOT gates (CX) is given for both approaches. The number of variational parameters is the same in both wave functions.}
  \label{tab:ressources_circuit}
\end{table}

\begin{figure}
  \centering
  \includegraphics[width = 0.8\textwidth]{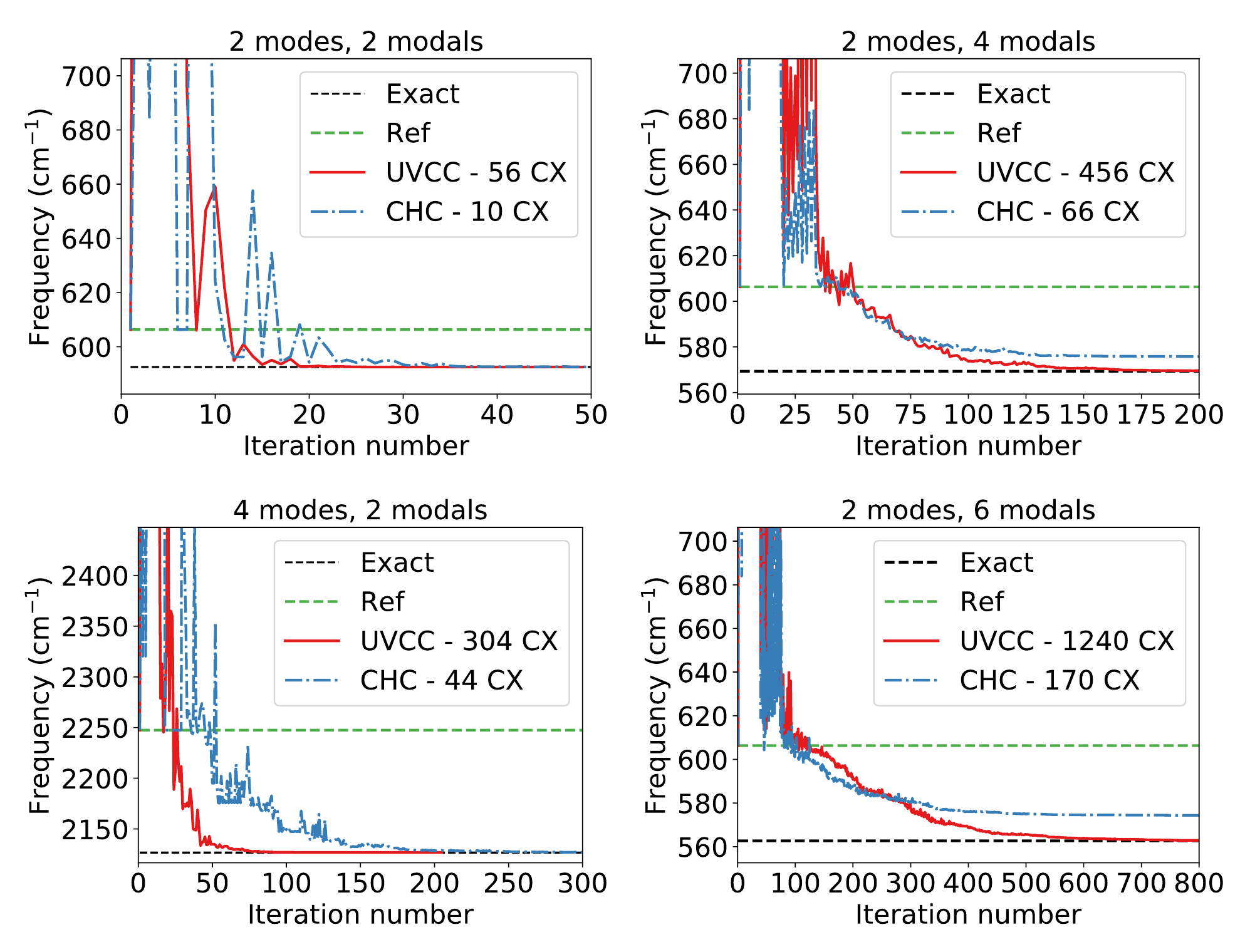}
  \caption{VQE convergence for CO$_2$ with different number of modes and modals based on the UVCC and CHC approaches including single and double excitations.}
  \label{fig:co2_w_chc}
\end{figure}

We study the CHC circuit for CO$_2$ with different number of modes and modals per mode (see Figure~\ref{fig:co2_w_chc}).
The results are of reasonable accuracy, with deviations $<15$ cm$^{-1}$ compared to the exact diagonalization energies.
As expected, CHC works best with two modals per mode since the entangling blocks are always applied to a state of the subspace with the correct symmetry, and therefore it does not create configurations with the wrong number of particles.
Errors increase with the number of modals for each mode.

\subsection{Quantum computation of vibrational structures on existing hardware}

\begin{figure}
  \centering
  \includegraphics[width = 0.5\textwidth]{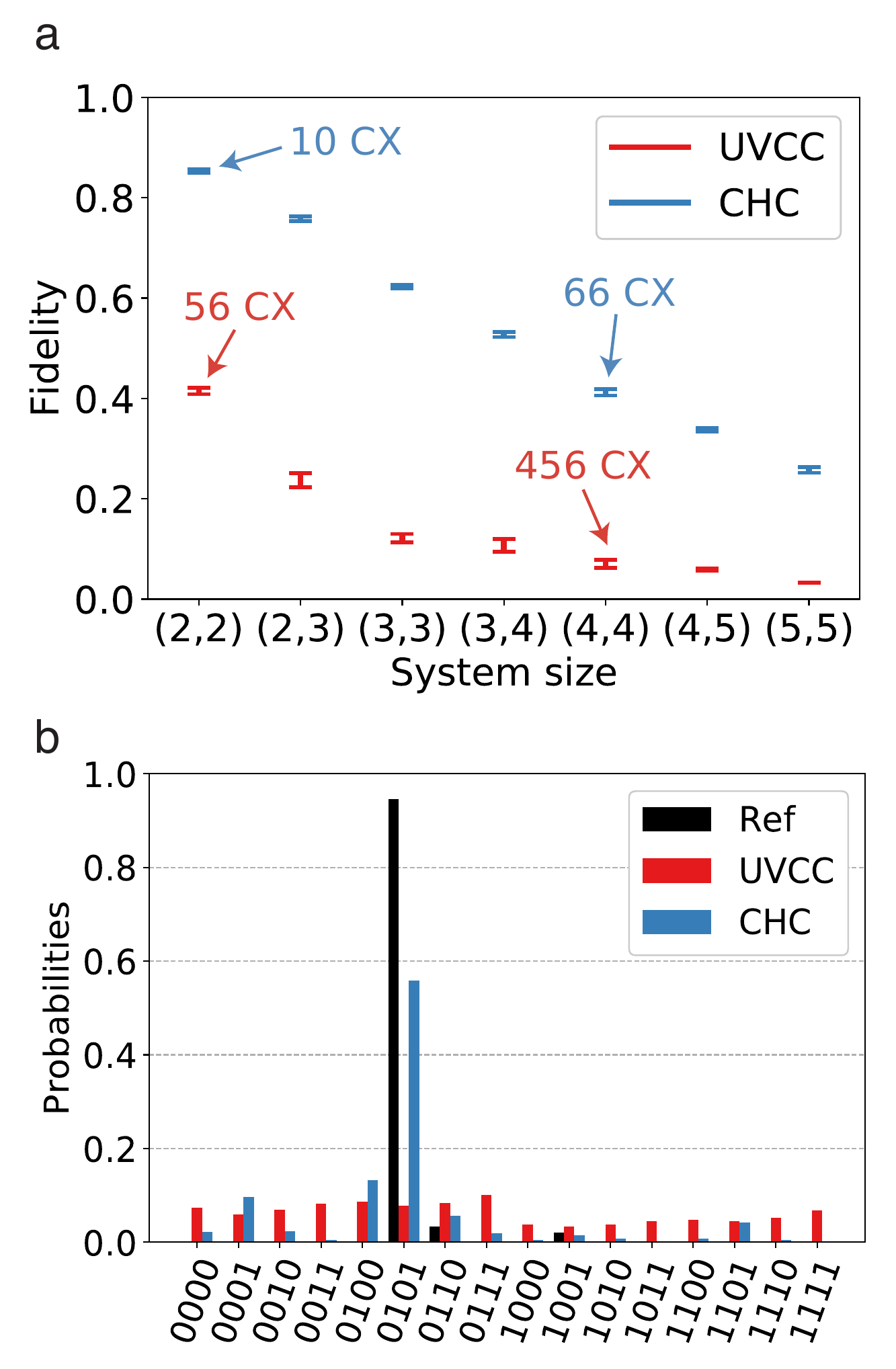}
  \caption{\textbf{a.} 
  Fidelities (according to Eq.~(\ref{eq:FidelityDef})) of the random state distributions computed for the first two vibrational modes of CO$_2$ obtained with the UVCC and CHC \textit{Ans\"atze}. The calculations are performed simulating the corresponding quantum circuits with noisy gate operations and for different number of modals for each of the 2 modes, as described by the pair ($N_{m_1}$, $N_{m_2}$). The fidelities are computed with 10000 measurements for each parametrized circuit. \textbf{b.} Histograms of the state probability distributions corresponding to the (2,2) setup obtained for both \textit{Ans\"atze} with the 20-qubit \textit{ibmq\_almaden} chip and 8000 measurements. We report the distributions for a single set of the qubit parameters. The reference corresponds to the exact solution obtained with the UVCC \textit{Ansatz}.}
  \label{fig:fidelity}
\end{figure}

In presence of typical hardware noise the loss in accuracy inherent to CHC will be balanced by the reduced circuit depth.
To study this effect, we use both UVCC and CHC in simulations with a noise model and in hardware calculations with the \textit{ibmq\_almaden} 20-qubit processor.
For this study, we fix the number of modes to two and progressively increase the number of modals per mode.
The gate angles, $\{\theta_i\}$, defining the circuit gates are initiated randomly and restricted to angles $-0.2\leq\theta_i\leq0.2$~\footnote{For small angles the reference configuration remains largely dominant. Hence, CHC is expected to lead to a reasonable approximation of UVCC for the same set of variational parameters.}.
The noise model includes only depolarization errors for single- and two-qubit gates, for which the error rate is based on the the average gate depolarization error associated to all qubits in the \textit{ibmq\_almaden} 20-qubit device.
The error rate values are $7 \times 10^{-4}$, $1.4 \times 10^{-3}$ and $2.2 \times 10^{-2}$ for U2, U3 and CNOT gates, respectively.
For each set of randomly parametrized UVCC and CHC circuits we evaluate the underlying probability distributions by performing 10000 measurements on the final states.
The final states are then compared to the reference obtained with the exact (noise-free) simulation of the UVCC \textit{Ansatz} (note that in both cases UVCC is the reference since we aim at approximating UVCC with CHC and assess the loss in accuracy due to the approximation versus the gain due to less noise).
This process is repeated 10 times renewing each time the parameter set. 
The fidelity of the resulting probability distribution is given by:

\begin{equation}
  \mathcal{F}^{\text{circ}} = 1 - \frac{\sum_i^{2^{N_q}}|C^{\text{circ}}_i-C^{\text{ref}}_i|}{\sum_i^{2^{N_q}}C^{\text{circ}}_i+C^{\text{ref}}_i}
  \label{eq:FidelityDef}
\end{equation}
where $C_i$ is the count for the occurrence of state $i$, with $i$ ranging over all possible $2^{N_q}$ states (${N_q}$ being the number of qubits).
The fidelities obtained with both UVCC and CHC are shown as a function of system size in Fig.~\ref{fig:fidelity}a.
The number of CNOTs is also given in Fig.~\ref{fig:fidelity}a for two modes and two modals per mode, and for four modes and four modals per mode.
These results prove that in the presence of noise a better accuracy is reached with CHC rather than with UVCC.
We repeat the experiment with a quantum processors (\textit{ibmq\_almaden}, 20 qubits) and a system size corresponding to two modes and two modals per mode (four qubits). 
Figure~\ref{fig:fidelity}b shows the histogram of the states distribution corresponding to a single trial (one set of gate parameters) confirming our observation that, with the present hardware, compact heuristic circuits such as the ones obtained with CHC outperform the original UVCC \textit{Ansatz}. 

\subsection{Quantum computation of the vibrational excited states of CO$_2$}

We calculate the excitation energies of a CO$_2$ molecule with (A) two modes, two modals per mode; (B) two modes, four modals per mode and (C) four modes, two modals per mode.
For all three cases, we limit the qEOM operators to 1- and 2-body excitation operators to restrict the number of measurements required to evaluate the matrix elements in Eq.~(\ref{eq:qEOM_MatrixElements}).
The ground state is approximated by running a VQE calculation with both UVCC and CHC.
The results of these (noise-free) simulations are presented in Table~\ref{tab:excited_states}.
The reference values are obtained from the exact diagonalization of the system Hamiltonian.

In case (A) all excitation energies are found with an accuracy $< 1$ cm$^{-1}$.
For (B), the accuracy is lower for CHC compared to UVCC. 
This is expected since for a large number of modals the accuracy of CHC ground state calculations decreases.
Finally, case (C) shows that higher order excitations need to be included in $\mathcal{E}_{\mu_\alpha}^{(\alpha)}$ in order to reach an accuracy of about 1 cm$^{-1}$ for the highest excitation energies with both wave function \textit{Ans\"atze}.

\begin{table}[htbp!]
    \centering
    \begin{tabular}{c c c c c c}
      \hline \hline
        & Modes & Modals & Reference & UVCC & CHC \\
      \hline
        A & 2 & 2 & \makecell{574.441\\1438.778\\2063.261} & \makecell{574.450\\1438.789\\2063.255} & \makecell{574.441\\1438.789\\2063.269} \\ 
      \hline
        B & 2 & 4 & \makecell{496.697\\1073.420\\1460.074\\1642.996\\2024.187\\2498.060} & \makecell{496.6680\\1073.418\\1460.084\\1642.978\\2024.123\\2498.037} & \makecell{479.2900\\1063.520\\1452.494\\1634.296\\2016.191\\2492.031} \\
      \hline
        C & 4 & 2 & \makecell{534.908\\559.330\\1098.527\\1267.081\\1855.895\\1880.816} & \makecell{534.682\\559.193\\1121.205\\1267.910\\1874.657\\1901.110} & \makecell{534.774\\559.274\\1121.298\\1268.086\\1874.732\\1901.130} \\
      \hline \hline
    \end{tabular}
    \caption{Lowest-lying excitation energies of a CO$_2$ molecule calculated with the qEOM algorithm and different circuit \textit{Ans\"atze}. The vibrational ground state is prepared with a classical simulation of the VQE algorithm.}
    \label{tab:excited_states}
\end{table}

\section{Discussion}

The proposed quantum algorithms for the calculation of the vibrational frequencies support PES representations and modals routinely employed in state-of-the-art traditional calculations. 
This enables a fair comparison of the scaling of the classical and quantum vibrational structure algorithms that provides an estimate of the resources required to reach a quantum advantage in vibrational-structure calculations.
As mentioned in the introduction, large-scale vibrational structure calculations are nowadays possible either with efficient VCI algorithms~\cite{Rauhut2009_VCI-Fast,Benoit2010_FastVCI,Rauhut2014_LiFClusters,Carrington2018_UracilNaphtalene,Bowman2019_EigenWater} or with non-linear wave function parametrizations, as is done in vDMRG~\cite{Baiardi2017_vDMRG} and VCC~\cite{Prasad1994_Original,christiansen2004,Christiansen2009_Automatic-VCC,Prasad2008_VCC-Response}.
The latter algorithm is the direct classical counterpart of our UVCC-based quantum algorithm and will be our reference for comparing the scaling with its quantum counterpart.

The scaling of VCC depends on the order of the highest degree of excitation that is included in the $\mathcal{T}$ operator (Eq.~(\ref{eq:ClusterOperator})).
For instance, VCC[2pt3] comprises all two-mode excitations and treats triple excitations perturbatively and can be applied, in its straightforward formulation, to molecules with up to seven atoms, such as ethylene oxide~\cite{Seidler2009_ThreeModeCoupling-VCC} including six modals for each vibrational mode.
The simulation of such molecules on quantum hardware would require 90 qubits.
The corresponding circuit that includes single and double excitations, and approximates triple excitations with the CHC \textit{Ansatz} (which we denote as UVCCSD(T)) would contain about $10^6$ CNOT gates.
By approximating also the single and double excitations with CHC, the number of 2-qubit gates drops to about $10^4$.
Currently, the state-of-the-art quantum hardware comprises about 50 qubits and has a coherence time supporting circuits with no more than $10^2$ CNOT gates.
The 2-qubit error rate of about $10^{-2}$ is currently the limiting factor for running such circuits in the state-of-the-art hardware.
By improving the 2-qubit gate fidelity and according to the estimated evolution of the quantum volume in superconducting
quantum computers~\cite{cross2019}, molecules of the dimensions of ethylene oxide will become accessible using the proposed algorithm within the next generation of quantum hardware. 

Molecules with up to seven atoms and described by a maximum of four-mode excitations can be studied with VCC by adopting a tensor-factorized representation of the amplitudes.\cite{Christianen2013_TensorDecomposition-VCC,Christiansen2018_VCC-TensorFactorized}
The inclusion of three- and four-body coupling terms in the potential and in the VCC wave function leaves the number of qubits required unchanged.
However, it also induces a raise in both the circuit depth and the number of measurements \textit{i.e.}, number of terms in the Hamiltonian.
For the electronic Hamiltonian, Motta and co-workers showed~\cite{Motta2018_LowRank} that tensor factorizations can be used to reduce the number of measurements and circuit depth for fermionic systems.
We expect that the same holds true also for the vibrational case, and we will consider such extension in future works.

The computational cost of VCC depends also on the choice of the coordinates used to describe the Hamiltonian.
It is known that for particular choices, such as local modes\cite{Reiher2009_LocalModes} or VSCF-optimized coordinates, the size of the off-diagonal anharmonic couplings can be significantly reduced.
This procedure has been already exploited to speed up traditional VCC calculations~\cite{Thomsen2014_OptimizedCoordinates-VCC} including only excitations between modes localized on the same portion of the molecules. 
This simplification has made VCC calculation feasible for systems as large as the water hexamer~\cite{Thomsen2014_OptimizedCoordinates-VCC}.
Our algorithm supports any choice of the reference coordinate system.
In the same way, the UVCC circuit can be adapted to a local mode representation by allowing only gates representing excitations for modes localized on nearby portions of the molecule.
For instance, in water clusters~\cite{Bowman2019_EigenWater} one could apply the UVCC circuit for excitations localized on one water molecule and include an approximated treatment of inter-fragment correlations with CHC.

\section{Conclusions}

We designed and compared different quantum computing strategies to calculate the vibrational structure of molecular systems amenable to near-term quantum computers.
We represented the vibrational wave function and PES in the $n$-mode-based Fock space~\cite{christiansen2004,Wang2009_SQMCTDH} that supports an arbitrary one-body reference basis and PES expressed as a generic many-body expansion.
This enabled us to overcome the limitations of recent algorithms~\cite{mcardle2019,sawaya2019} that rely on a harmonic-based reference and are, therefore, not flexible enough for strongly anharmonic systems.
We compared state-of-the-art circuits to prepare the wave function of a two-dimensional Hamiltonian modelling the nuclear dynamics of CO$_2$ and introduced the Compact Heuristic circuit for Chemistry (CHC).
On the one hand, the Unitary Vibrational Coupled Cluster (UVCC) delivers the most accurate vibrational energies, but at the price of very deep circuits that are difficult to implement on currently available quantum hardware already for three-atom molecules.
On the other hand, heuristic circuits provide less accurate vibrational energies, while being shallower.
In this work, we showed how CHC represents an optimal compromise combining the advantages of UVCC and heuristic wave function approaches.
However, the CHC wave function does not fulfill the symmetries of the vibrational Hamiltonian.
Therefore one first needs to project it onto the correct symmetry subspace before evaluating the vibrational energy.
This effect is only minor for the systems studied here, but we expect it becomes larger for strongly anharmonic molecules.
In those cases, a modal basis obtained from a VSCF calculation can significantly improve the accuracy.
A second limitation of our algorithm is that each modal is mapped to a different qubit.
Therefore, a large portion of the qubit Hilbert space does not correspond to a physically acceptable state.
Occupation number vectors based on alternative mappings, such as the ones introduced in Ref.~\citenum{sawaya2019}, produce more compact representations of the vibrational states.
However, the reduction in the number of qubits comes at the cost of an increase in the circuit depth for the representation of the wave function.
Heuristic circuits inspired by the CHC strategy, but adapted to these more compact mappings, could enable the calculation of vibrational energies for molecules with more than three atoms on a state-of-the-art quantum computer.
Finally, we extended the quantum Equation of Motion (qEOM)\cite{ollitrault2019} approach to calculate vibrational excitation energies and applied it to the CO$_2$ molecule.

This work also sets the fundamentals for the quantum computation of the ground-state energy of interacting fermions and bosons, such as polaronic~\cite{Macridin2019_PolaronModel} or quantum optics Hamiltonians~\cite{DiPaolo2019_VQE-LightMatter}.
In the quantum-chemistry context, this is the case of the pre-Born-Oppenheimer molecular Hamiltonian~\cite{Bubin2013_Review,Reiher2012_ECG} that has been studied so far with the quantum phase estimation algorithm.\cite{Veis2016_QC-PreBO}
Moreover, the algorithm can be extended to the time domain (e.g., using the time-dependent Schr\"{o}dinger formalism of Ref.~\citenum{Benjamin2017_VQE-Dynamics}) to address quantum dynamics.

\begin{acknowledgement}
This work was supported by ETH Z\"{u}rich through the ETH Fellowship No.~FEL-49 18-1.\\
The authors acknowledge financial support from the Swiss National Science Foundation (SNF) through the grant No. 200021-179312.\\
IBM, IBM Quantum, \texttt{Qiskit} are trademarks of International Business Machines Corporation, registered in many jurisdictions worldwide. Other product or service names may be trademarks or service marks of IBM or other companies.
\end{acknowledgement}

\providecommand{\latin}[1]{#1}
\makeatletter
\providecommand{\doi}
  {\begingroup\let\do\@makeother\dospecials
  \catcode`\{=1 \catcode`\}=2 \doi@aux}
\providecommand{\doi@aux}[1]{\endgroup\texttt{#1}}
\makeatother
\providecommand*\mcitethebibliography{\thebibliography}
\csname @ifundefined\endcsname{endmcitethebibliography}
  {\let\endmcitethebibliography\endthebibliography}{}

\end{document}